# Optimal conditions and the dynamics of quantum memory for spatial frequency grating resonators

N. S. Perminov[a, b], K. V. Petrovnin[c], R. S. Kirillov[c], R. R. Latypov[c], S. A. Moiseev[a, b, *], and O. N. Sherstyukov[c]

[a]*Kazan Quantum Center, Kazan National Research Technical University, Kazan, 420111 Russia*

[b]*Zavoisky Physical-Technical Institute, Russian Academy of Sciences, Kazan, 420029 Russia*

[c]*Kazan (Volga Region) Federal University, Kazan, 420000 Russia*

*e-mail: s.a.moiseev@kazanqc.org

**Abstract.** The dynamics of the interaction between microcavities connected to a common waveguide in a multiresonator quantum memory circuit is investigated. Optimum conditions are identified for the use of quantum memory and a dynamic picture of the exchange of energy between different microcavities is obtained.

## INTRODUCTION

Let us discuss the dynamics of the interaction between quantum subsystems in a quantum memory circuit that is integrated into a waveguide system and is capable of retaining a wideband microwave signal over much longer times than the duration of the microwave field pulses. The initial idea is based on approaching a quantum memory as a photon echo [1] of the type found in atomic systems with a periodic spectral structure of the inhomogeneous resonant transition broadening [2] known as an AFC protocol.

In the considered multiresonator scheme (the MR scheme shown in Fig. 1), instead of an atomic system we use a system of several microresonators connected to a common waveguide and arranged at regular intervals in the waveguide, producing a periodic discrete structure of narrow resonant lines as well. An experimental prototype of such a resonator system for the microwave frequency range is shown in Fig. 2. The setup includes a section of the microwave waveguide that has a wide transparency band for the microwave signal field. It is capable of transmitting any information that has to be retained in the microresonator system for periods of time much longer than the duration of an incoming pulse.

To retain the information of the signal field, we connected microresonators to the lateral walls of the microwave waveguide; these were cylindrical resonators made of a composite material and containing a dielectric with a high refractive index. Because of this, the diameter of the microresonators could be made smaller than the wavelength of microwave radiation in the waveguide. This allows us to position a sufficient number of the microresonators along the waveguide.

The electromagnetic field of every microresonator interacts with the microwave field in the waveguide through a narrow slit in their common wall. Several microresonators can be arranged along the waveguide wall at distances close to the radiation wavelength, resulting in the virtually simultaneous absorption of a signal pulse by the entire system of microresonators. Our numerical calculations of the MR scheme and the initial experiments confirmed the possibility of stably positioning of several resonant lines of microresonators near one another, even when they are in close proximity inside the waveguide.

## THEORETICAL MODEL

In theoretical simulations of the interaction between the resonators and the electromagnetic field

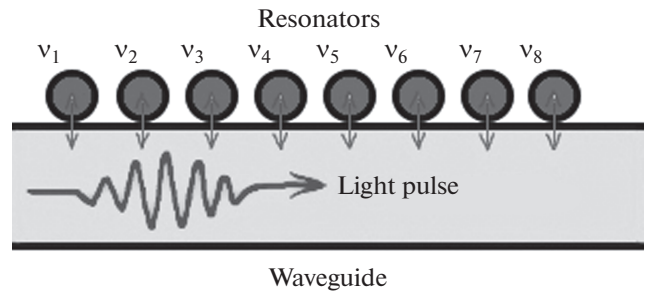

**Fig. 1.** MR scheme of quantum memory. The waveguide is coupled to a system of resonators through slits in the wall. The distance between neighboring resonators is equal to the wavelength, and the frequencies of the resonators form a periodic structure. The frequency and $Q$-factor of each resonator can be controlled independently.

in the waveguide, we used the Jaynes–Cummings Hamiltonian [3]

$$\hat{H} = \int dk c|k| \hat{b}_k^+ \hat{b}_k + \sum_n [ck_0 + \Delta n] \hat{a}_n^+ \hat{a}_n \\ + g \int dk \sum_n \left[ e^{-izkn} \hat{b}_k^+ \hat{a}_n + \text{H.C.} \right], \quad (1)$$

where $c$ is the speed of light; continuous index $k \in [-k_0 + \delta_0; -k_0 - \delta_0] \cup [k_0 + \delta_0; k_0 - \delta_0]$ corresponds to direct reverse waves in a waveguide with carrier frequency $ck_0$; $\delta_0$ is the waveguide band width; discrete index $n \in \{1, ..., N\}$ is used to number the resonators; and $\hat{a}_n^+, \hat{b}_k^+$ and $\hat{a}_n, \hat{b}_k$ are the boson operators of the creation and annihilation of photons, respectively. The first summand in (1) corresponds to the energy of the field in the waveguide; the second, to energy $N$ of microresonators with frequencies $ck_0 + \Delta n$ (where $\Delta$ is the frequency offset); and the third, to the point interaction between the waveguide field and the field of microresonator, which is controlled using dimensional interaction constant $g$ and distance $z$ between neighboring resonators.

With interaction between the system of microresonators and the single-photon signal field at the fairly low initial temperature of the microresonators and the waveguide, relative to the energy of a radiation quantum, the wave function of the entire quantum system can be presented as the direct sum of quantum subsystems of the form:

$$|\psi\rangle = \left[ \int dk f_k(t) \hat{b}_k^+ + \sum_n \alpha_n(t) \hat{a}_n^+ \right] |0\rangle, \quad (2)$$

where $f_k(t)$ are the electromagnetic field modes in the waveguide, and $\alpha_n(t)$ are the electromagnetic field modes in the resonators. For a single-photon field, the condition of wavefunction normalization is written in the form $\int dk f_k^*(t) f_k(t) + \sum_n \alpha_n^*(t) \alpha_n(t) = 1$.

We are interested in obtaining a highly efficient memory, for which the system's energy can be concentrated in the microresonators in the middle of the storage phase. We thus describe this system's dynamics starting from the reemission phase ($f_k(0) = 0$, $\alpha_n(0) = c_n$). This approach offers considerable computational advantages when determining the conditions for obtaining a highly efficient quantum memory. Note that the convenient analytical form of the solutions obtained in this way allows highly accurate control of the quantum memory's dynamic parameters.

The Schrödinger equation $[\partial_t + i\hat{H}]|\psi\rangle = 0$ leads us to a system of integro-differential equations for the amplitudes of the electromagnetic field modes in the waveguide and the microresonators:

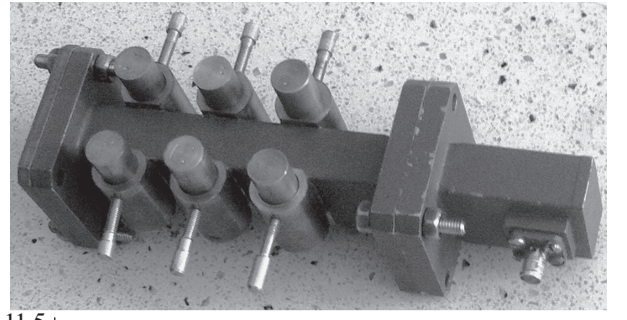

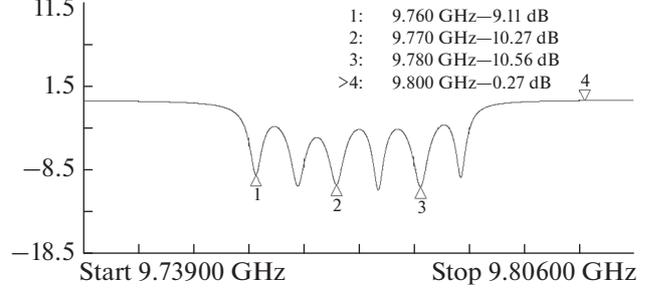

**Fig. 2.** Experimental prototype and its spectrum: periodic structure of spectral lines of resonators (carrier frequency, ~10 GHz; distance between peaks, ~5 MHz) showing that quantum memory can be achieved experimentally.

$$[\partial_t + ic|k|] f_k(t) + ig \sum_n e^{-ikzn} \alpha_n(t) = 0, \\ f_k(0) = f_k^0, \quad [\partial_t + i\Delta n] \alpha_n(t) + ig \int dk e^{+ikzn} \\ \times f_k(t) = 0, \quad \alpha_n(0) = \alpha_n^0, \quad (3)$$

where $f_k^0 = 0$, $\alpha_n^0 = c_n$, which corresponds to the problem of electromagnetic field radiation from resonators into a common waveguide. The normalization conditions can then be rewritten as $\sum_n |c_n|^2 = 1$.

The formal solution for the electromagnetic field modes in the waveguide can in this case be expressed through the electromagnetic field modes in the microresonators, using the formula

$$f_k(t) = -ig \int_0^t d\tau \sum_m \exp^{(-i|k|[c(t-\tau) + \text{sign}(k)zm])} \alpha_n(\tau), \quad (4) \\ f_k(0) = 0, \quad \alpha_n(0) = c_n,$$

where $\text{sign}(k)$ is the wave vector sign. It should be noted that the electromagnetic field amplitudes in the waveguide corresponding to different directions of wave propagation are generally not of the same magnitude for identical fields $\alpha_n(\tau)$ in the microresonators through which they are determined. This means we can in principle use MR schemes to transform the signal field spectrum. From a physical standpoint, this is an indication that some forms of the signal field can alter the parameters of the spectrum when stored. This is a fundamental attribute of all AFC protocol variations and imposes a spectral matching requirement [4–6]. In our case, the spectral matching requirement is weak and corresponds to $c_n^* = c_{-n}$, which imposes no appreciable limitations on the form of the signal field. This matter will be dealt with in more detail in our next works.

After extracting rapidly oscillating part $\alpha_n(t) = \exp^{(-ik_0ct)} \beta_n(t)$ from Eq. (3), we get the system of equations

$$[\partial_t + i\Delta n]\beta_n(t) + \frac{\pi g^2}{c}\sum_m \beta_m(t) = 0, \quad \beta_n(0) = c_n, \quad (5)$$

where we allow for the distance between resonators being equal to wavelength $z = 2\pi/k_0$, and the field pulse duration is much longer than time $\delta t_{\mathrm{signal}} \gg (n-m)z/c$ in which the signal passes the waveguide.

## ANALYTICAL SOLUTIONS AND OPTIMUM CONDITIONS

The solution to Eqs. (5) has the form

$$\beta_n(t) = \exp^{(-i\Delta nt)} \left[ c_n + \frac{\pi g^2}{c\Delta}\left[1 + \frac{\pi^2 g^2}{c\Delta}\right]^{-1} \right. \\ \left. \times \sum_m \frac{\sin(\Delta(n-m)t/2)}{(n-m)t/2} e^{i\Delta(n-m)t/2} c_m \right]. \quad (6)$$

Numerical calculations of the initial equations show that formula (6) is highly accurate even for a small number $N$ of microresonators (its precision is 0.1% even when $N = 6$). From formula (6) we find the quantum efficiency of field restoration $\eta(t) = 1 - \sum_n |\alpha_n(t)|^2$, which takes the form of the product

$$\eta(t) = \eta_0(g)\eta_1(t), \quad \eta_0(g) = \frac{4\pi^2 g^2}{c\Delta}\left[1 + \frac{\pi^2 g^2}{c\Delta}\right]^{-2}, \\ \eta_1(t) = \int_0^{\Delta t/2\pi} d\tau \left|\sum_n \exp^{(i\Delta n\tau)} c_n\right|^2. \quad (7)$$

An important conclusion follows from formulas (7) for the quantum efficiency: 100% efficiency can be achieved for finite optimum values of the interaction parameters. Indeed, the maximum of function $\eta_0(g)$ is equal to unity and can be achieved if a single matching condition is met:

$$\pi^2 g^2 = c\Delta. \quad (8)$$

The dynamic picture of the quantum efficiency under optimum conditions is determined by function $\eta_1(t)$. Since re-normalization condition $\sum_n |c_n|^2 = 1$ is met, $\eta_1(2\pi/\Delta) = 1 \forall c_n$ for any initial data $c_n$ in the re-emission problem. The MR scheme of quantum memory is thus a universal one.

Note that for the AFC protocol, or for any other version of quantum memory based on a photon echo, in which an atomic ensemble in a single-mode resonator is used to achieve 100% efficiency, it is required that two matching conditions be met, as was demonstrated in [4]. In the considered situation, it is sufficient to meet only one condition, (8). Experimental implementation of the broadband quantum memory is greatly facilitated by the described properties of the optimality conditions. Below, we shall assume that this condition is met. We stress, however, that changes in dimensional coupling constant $g$ can also be used to control the quantum entanglement of microresonators states. This matter requires special study and will be considered in more detail in our subsequent works.

## DYNAMICS OF MICRORESONATOR INTERACTION

Interaction between microresonators in the MR scheme of quantum memory displays rich dynamics described by the general solution to (6). As regards applications, one of important problem of the dynamics of such a system is the problem of energy exchange between microresonators during their interaction. This is explained by the resolving ability of the ways of registering dynamic parameters of quantum systems being often linked to the energy and intensity of excitation of interacting subsystems, with which detectors of the incoming radiation can in turn selectively interact. It is therefore of interest to determine the comparative energy characteristics of two resonators to which detectors can be connected independently. In such a configuration, detector sensitivity can be improved considerably by using the drastic dynamic picture of the energy exchange between subsystems (microresonators).

To explore the exchange pattern, we introduce the relative energy difference of interacting resonators 1 and 2 according to the formula

$$e_{12}(t) = \frac{E_1(t) - E_2(t)}{E_1(t) + E_2(t)}, \quad (9)$$

where $E_k(t) = |\beta_{n_k}|^2$ is the energy of the resonator with number $n_k$ ($k \in \{1, 2\}$). Note that this quantity is fairly stable with respect to noise in the waveguide because it contains the energy difference, while the contribution from noise to the energy of neighboring resonators having similar energy amplitudes is virtually identical over brief time intervals.

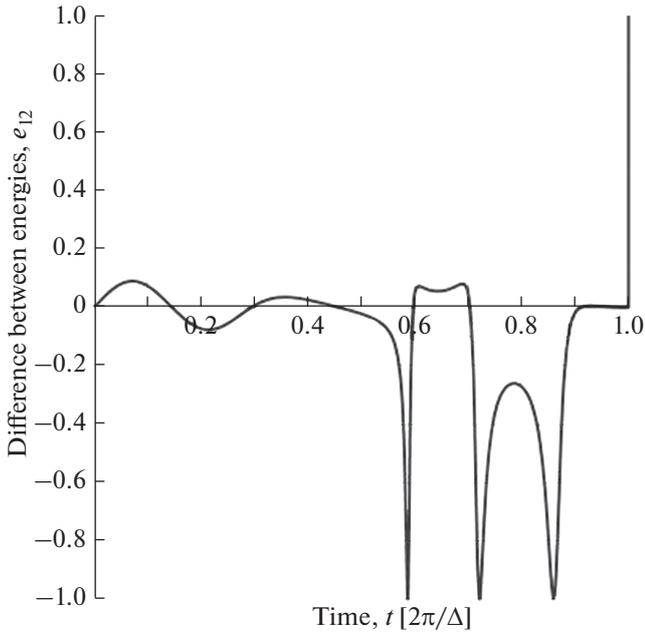

**Fig. 3.** Relative difference in the energies between the two resonators in the center when $N = 7$. Three sharp peaks much narrower than pulse duration $2\pi/(N\Delta)$ are seen, detailing the pattern of quantum entanglement.

Let us consider the important case of equal initial field amplitudes in the microresonators, corresponding to a short pulse of rectangular form: $c_n = (-1)^n / \sqrt{N}$. The relative energy difference between the central resonator and its neighbor is shown in Fig. 3 for $N = 7$. The obtained graph has three peaks whose width is much less than pulse duration $2\pi/(N\Delta)$. The shape and position of these short peaks is very sensitive to system parameters and external factors. At certain moments in time, large amounts of energy can thus be concentrated in only one of two microresonators located close to one another. Owing to such concentrations of energy in one of the microresonators at certain moments, the electromagnetic field's interaction with a detector bound to such a highly excited resonator can be improved substantially, which also permits more detailed study of the dynamics of resonators using experimental means.

Note that as the process evolves, the system of microresonators can transition to a state with a high degree of quantum entanglement. This is accompanied by the emergence of so-called *dark* and *bright* quantum states, which allows us to store incoming data for a specified time and retrieve it again. It is the effective buildup of bright states in the system of microresonators that makes the high quantum efficiency of the considered MR scheme memory possible. It is therefore of interest to search for additional ways of controlling the parameters of microresonators in order to increase storage time and control their quantum states as we desire.

## CONCLUSIONS

A model of broadband quantum memory based on a system of microresonators without the use of an atomic ensemble was built. In the studied system, the possibility of independently controlling separate subsystems (microresonators) was demonstrated, allowing changes in the spectral characteristics of the scheme for storing and retrieving quantum information. An experimental prototype for the microwave range was built, on which basic spectral properties of the MR scheme were successfully tested.

Optimum conditions were identified for implementation of the MR scheme, which can be achieved in actual experiment. The possibility of achieving 100% efficiency of quantum memory was demonstrated for the optimum configuration of controlling parameters, and the universal character of the MR scheme was proved.

The theoretical possibility of resolving the temporal dynamics of individual microresonators over brief time intervals of their interaction with the waveguide field was demonstrated. This observation is of great interest when choosing the technique for optimum detection of weak external radiation. In addition, we can study in detail the quantum entanglement necessary for experimental implementation of microwave quantum computer architecture.

A unique feature of the considered scheme is the possibility of the precise experimental allocation of parameters of composite resonators, making it easy to achieve the temporal dynamics of the required profile and create on its basis a scheme for the synchronization of nonclassical elements that can be integrated into a quantum computer.

## ACKNOWLEDGMENTS

This work was funded under the State Program for Improving the Competitiveness of Kazan (Volga Region) Federal University as a Leading Scientific Center; and by the Russian Foundation for Basic Research, grant no. 15-42-02462/16.